\documentclass[runningheads]{llncs}
\usepackage[T1]{fontenc}
\usepackage{graphicx}
\usepackage{amsmath}
\usepackage{amssymb}
\usepackage{booktabs}
\usepackage{url}
\usepackage{float}
\usepackage{xcolor}
\usepackage{tikz}
\usetikzlibrary{arrows.meta,calc,fit,positioning}
\begin{document}
\title{PyDoseRT Proton: A GPU Pencil-Beam Engine with a
Convolutional Residual-Correction Network for Fast Proton Dose Calculation}
\titlerunning{PyDoseRT Proton}
\author{Lukas Zimmermann\inst{1,2} \and
Hermann Fuchs\inst{1,2} \and
Attila Simk\'o\inst{3} \and
Gerd Heilemann\inst{1,2}}
\authorrunning{L. Zimmermann et al.}
\institute{Department of Radiation Oncology, Medical University of Vienna,
Vienna, Austria \and
Christian Doppler Laboratory for Image and Knowledge Driven Precision
Radiation Oncology, Medical University of Vienna, Vienna, Austria \and
Department of Diagnostics and Intervention, Ume\aa{} University, Ume\aa{}, Sweden}
\maketitle
\begin{abstract}
\textbf{Architecture category.} Hybrid method: a physics-based
analytical pencil-beam (PB) dose engine followed by a 3-D convolutional
residual-correction network (RepVGG--U-Net).

We addressed the DoseRAD2026 proton dose-prediction task with \emph{PyDoseRT
Proton}, a GPU-accelerated engine implemented in PyTorch and augmented by a
learned residual toward Monte Carlo (MC) accuracy. A
double-Gaussian PB kernel was calibrated to GATE/Geant4 integrated depth doses
in water in two stages: a classical per-energy curve fit, then a gradient-based
fit of the full 3-D dose through the PyTorch physics engine as it retains a differentiable execution path for
gradient-based optimization of dose-dependent objectives. The engine computes each beamlet on a beam's-eye-view (BEV) lattice with variance-preserving
Gaussian splitting, an analytic nuclear halo, and a Fermi--Eyges heterogeneity
term, then rotates the result into the patient frame. Additionally, a compact residual U-Net predicts an \emph{additive} correction in BEV
space. It is conditioned on voxelwise material-label embeddings, a discrete
energy embedding and spot size. 
The same model was used for all anatomical sites (thoracic and abdominal). It was
trained with a patient-space $L_1$ objective emphasizing the scored high-dose
region and multi-scale BEV deep supervision. The submitted CT configuration
obtained preliminary-test beamlet MAE $0.0066$, image-$z$ IDD distance $0.0025$,
plan MAE $0.0049$, $98.30\%$ gamma pass rate (1\%/1\,mm), and DVH error $0.460$.
\end{abstract}
\section{Introduction}
Intensity-modulated proton therapy delivers highly conformal dose but is
sensitive to range uncertainty and anatomical changes, which motivates dose
engines that are both fast enough for online adaptation and accurate enough to
replace analytical pencil-beam (PB) algorithms near tissue heterogeneities.
Monte Carlo (MC) transport is the accuracy reference but is too slow for
iterative planning; analytical PB algorithms~\cite{ref_hong} are fast but degrade in lung and
at bone/air interfaces. Recent work showed that a \emph{differentiable} dose
engine enables gradient-based plan optimization end to end~\cite{ref_pydose}.

We extend this idea to protons for the DoseRAD2026 challenge. Our method couples
a proton PB engine, calibrated to MC in water through an autograd-enabled copy
of its forward model, with a compact convolutional network that learns an
additive residual toward MC. The zero-initialized residual anchors optimization
at the analytical solution while retaining an explicit physics baseline instead
of predicting dose directly from anatomy.

\section{Methods}
\subsection{Data}
\label{sec:data}
We used the DoseRAD2026 proton dataset~\cite{ref_doserad}. The dose corrector was trained on $67$
cases and hyperparameters were tuned on the $8$ held-out validation cases
(four abdominal: 1ABB006/011/020/021; four thoracic: 1THB002/008/011/016);
the training cohort contained 32 abdominal and 35 thoracic cases. The learning target was the provided
MC reference dose. Inputs were the planning CT, from which energy-dependent
stopping-power ratio (SPR), material labels, and physical mass density were
derived, and the plan geometry. The synthetic-CT (sCT) model was trained separately using all 75 paired MR/CT volumes. No additional patient data were used.

For \emph{kernel calibration only} (not network training) we generated
additional reference data outside the challenge set: OpenGATE v10.1 alongside Geant4~v11.4 ~\cite{ref_gate} was used to simulate proton beams in a homogeneous water
phantom ($30\times10\times10\,\mathrm{cm}^3$, $0.2\,\mathrm{mm}$ isotropic
resolution, $1500\times500\times500$ voxels). Physics list ($\mathrm{QGSP\_BIC\_EMV}$), production cuts and step sizes, initial energies as well as energy spread were selected based on the challenge MC simulation template and base data. Successive data sets used
$10^7$, $10^8$, or $10^9$ primary histories per energy. The production
$10^8$ set contains 85 energies from $31.7$ to $200.8\,\mathrm{MeV}$; the
earlier $10^7$ fit contains 43 calibrated energies, whereas the matched
$10^9$ refit uses the same 85 energies as $10^8$. The shipped system uses the $10^8$ LUT.

\begin{figure}[htpb!]
\centering
\includegraphics[angle=270,width=\textwidth,  trim=3cm 0cm 3cm 0cm,
  clip]{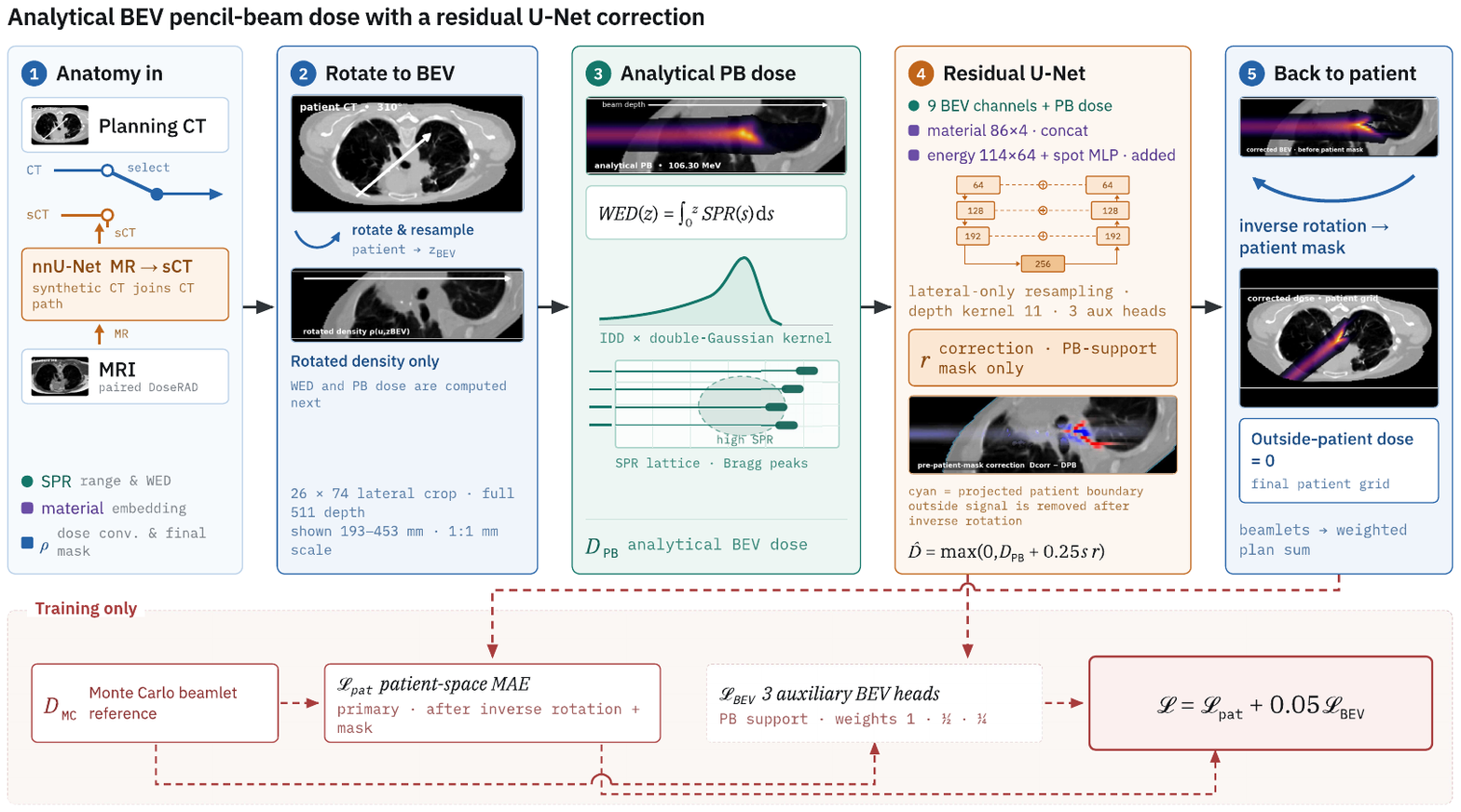}
\caption{PyDoseRT Proton workflow. Patient data are rotated into the beamlet's
BEV frame for analytical dose calculation and learned correction, then mapped
back to the patient grid. Training combines the official patient-space target
with multi-scale BEV deep supervision; material, energy, and spot size condition
the corrector through spatial and global pathways.}
\label{fig:workflow}
\end{figure}

\subsection{Model}
\label{sec:model}

\paragraph{Overall architecture.}
The proposed framework has two input-specific pathways that share the same proton dose-calculation backend (see Fig 1). In the CT pathway, the planning CT was converted directly into physical mass density, discrete material labels, and energy-dependent stopping power ratio (SPR) maps. In the MRI pathway, an independently trained nnU-Net regression model first generated a sCT on the common CT grid, from which the same physical input maps were derived. All subsequent processing was identical for both pathways: the patient data were transformed into the beam’s-eye-view coordinate system, dose was calculated using the analytical pencil-beam engine, and the residual-correction network predicted an additive correction toward the Monte Carlo reference. The corrected dose was then transformed back to the patient grid and post-processed. The components are described below.

\paragraph{Pencil-beam dose engine.}
The pencil beam dose engine was based upon a pencil beam dose algorithm developed in our group \cite{ref_Fuchs2012}. Each spot was a beamlet processed on its BEV lattice. Along every lattice column
the water-equivalent depth is
$d(k)=\Delta\sum_{j\le k}S_j-\tfrac12\Delta S_k$, where $S_j$ is SPR
(voxel-center convention), so the Bragg peak shifts laterally and longitudinally with heterogeneity.
Each beamlet is split into $9\times9=81$ sub-beams at quarter-FWHM spacing with a
variance-preserving envelope ($\sigma_\mathrm{sub}=\sigma_\mathrm{spot}/\sqrt n$),
which approximates the incident Gaussian while retaining its variance. The narrow core
(Eq.~\ref{eq:double}) is deposited per sub-beam; the broad halo is added once per
beamlet. 
A Fermi--Eyges/Kanematsu term~\cite{ref_kanematsu} models beam broadening due to multiple Coulomb scattering.

Dose is rotated into the patient
frame by a matched bilinear resampling.
For efficiency, challenge inference ran without recording an autograd graph.
However, the engine's PyTorch operation stack could retain an autograd-enabled execution path. 
In this study we used that
capability for Stage-2 LUT calibration with respect to the kernel curves. However, gradients can be even propagated through the dose calculation to optimize treatment-plan parameters, enabling gradient-based calibration and, in principle, plan optimization against dose objectives. 

\paragraph{Kernel and water calibration.}
The lateral kernel is a radially symmetric, separable double Gaussian,
integrated analytically over each voxel cell,
\begin{align}
K(x,y)&=(1-w)\,g(x;\sigma_1)\,g(y;\sigma_1)+w\,g(x;\sigma_2)\,g(y;\sigma_2),
\label{eq:double}\\
g(x;\sigma,h)&=\tfrac12\!\left[\operatorname{erf}\tfrac{x+h}{\sqrt2\,\sigma}
-\operatorname{erf}\tfrac{x-h}{\sqrt2\,\sigma}\right],\notag
\end{align}
with narrow core $\sigma_1$, broad nuclear halo $\sigma_2$ and halo weight $w$,
all depth-dependent; a per-energy IDD table $Z(d)$ supplies the amplitude. It is
calibrated to the MC water data in two stages: (i)~a direct normalized
double-Gaussian fit of $(\sigma_1,\sigma_2,w)$ and $Z$ at each depth using
trust-region least squares; and (ii)~a gradient-based refinement of all 
depth curves by back-propagating the normalized 3-D water dose through an
autograd-enabled calibration path (Adam, learning rate $0.03$, cosine schedule,
150 iterations per energy, $0.4\,\mathrm{mm}$ depth sampling). The calibration
loss is masked below $0.5\%$ of the MC peak and includes a second-difference
smoothness prior. 

\paragraph{Correction network.}
A 3-D convolutional residual U-Net (\emph{RepVGG--U-Net}~\cite{ref_repvgg}, 887,292 trainable
parameters) operates per beamlet in BEV space. The v1 input comprises eight
continuous channels:
SPR minus one, lateral radius, normalized geometric depth, normalized WED, two
signed lateral offsets, non-negative SPR, and peak-normalized PB dose. V2 differs
architecturally only by appending a ninth, range-relative channel
$\delta_R=(\mathrm{WED}-R_{\mathrm{peak}}(E))/100\,\mathrm{mm}$; the engine,
conditioning paths, and U-Net topology are unchanged. This widens only the
input projection, increasing the parameter count from 887,228 to 887,292. In
addition, each voxel's
discrete material label is mapped to a learned four-dimensional embedding and
concatenated with the continuous feature volumes before the input projection.
For each beamlet, the scalar energy is assigned to the nearest of 114 tabulated
machine energies and mapped to a learned 64-dimensional class embedding. A two-layer
$2\!\rightarrow\!64\!\rightarrow\!64$ multilayer perceptron (MLP) encodes the two lateral spot widths;
energy and spot-size vectors are summed, broadcast spatially, and added to
the 64-channel native feature map at the U-Net entrance. Thus material
conditioning is spatial, whereas energy and spot-size conditioning is global.

This conditioning design was selected in an early controlled screen using a
smaller U-Net with native width eight. We compared the learned energy-class
embedding with no energy input and with either a normalized scalar or nine
Fourier features (the scalar plus four sine/cosine pairs), projected by
$1\!\rightarrow\!8\!\rightarrow\!8$ and $9\!\rightarrow\!8\!\rightarrow\!8$
SiLU MLPs, respectively. Spot widths used a
$2\!\rightarrow\!8\!\rightarrow\!8$ SiLU MLP. Spot-width conditioning and
GroupNorm versus InstanceNorm were varied in the same matrix. The selected mechanism was then
scaled with the native width from eight to 64; the four-dimensional material
embedding was held fixed and was not independently swept. We also compared
adding the global conditioning vector at the U-Net entrance with adaptive
GroupNorm modulation (AdaGN).

Down/up-sampling is applied only laterally; full beam-axis depth resolution is
retained. The encoder/deep path uses widths 64, 96, 128, 192, and 256 with
2, 2, 2, 4, and 4 RepVGG blocks, respectively. An initial factor-three
reduction between the 64- and 96-channel stages equalizes the anisotropic crop,
followed by three factor-two reductions. The decoder is intentionally
shallower: it applies two RepVGG blocks after each up-projection to 192, 128,
and 96 channels, then projects directly to the native 64-channel resolution
without another RepVGG stage. Skip connections are additive, and auxiliary
corrected-dose heads are attached to the 192-, 128-, and 96-channel decoder
stages. RepVGG blocks combine a length-11 depth convolution and a $3\times3$
lateral convolution with eight-group GroupNorm and SiLU; their train-time
branches are folded for inference. The native head predicts an additive
residual,
\begin{equation}
\hat{D}=\max\!\big(0,\;D_\mathrm{PB}+s\cdot\alpha\cdot r\big),\qquad\alpha=0.25,
\label{eq:residual}
\end{equation}
where $s$ is the per-beamlet PB peak. All residual-output heads are
zero-initialized, so the network begins at the PB prediction.
No externally pre-trained weights are used for the dose corrector.

\paragraph{MRI-to-CT conversion.}
The MRI pathway uses a 3-D residual-encoder nnU-Net regression model~\cite{ref_nnunet} to generate a synthetic CT from each MRI. The model follows the \texttt{nnUNetResEncModelM} configuration and comprises six encoder stages with 32, 64, 128, 256, 320, and 320 channels. Its output is defined on the common CT grid and provides the input to the same physical-map construction, analytical pencil-beam engine, and residual-correction network used in the CT pathway.

\paragraph{Post-processing.}
The corrected BEV dose was rotated back and densified onto the patient grid and
clamped to nonnegative values. The patient support mask was topological: the
largest connected external-air component was removed and the per-slice body
contour was filled, preserving internal air cavities such as bowel gas and the
trachea.

\subsection{Training}
Let $e=|\hat D-D_\mathrm{MC}|/\max(D_\mathrm{MC})$ for one beamlet and let
$\langle\cdot\rangle_A$ denote the mean over mask $A$. The primary loss was
computed after inverse rotation in patient space,
\begin{equation}
 \mathcal L_\mathrm{pat}=\langle e\rangle_{D_\mathrm{MC}\ne0}
 +0.15\,\langle e\rangle_{D_\mathrm{MC}>0.1\max(D_\mathrm{MC})}.
 \label{eq:patientloss}
\end{equation}
During training, the MC reference was also sampled into BEV. Three auxiliary
decoder predictions were compared with average-pooled BEV targets using masked,
peak-normalized $L_1$ losses $\ell_j$. Their relative weights halve with scale,
\begin{equation}
 \mathcal L=\mathcal L_\mathrm{pat}+0.05\,\mathcal L_\mathrm{BEV},\qquad
 \mathcal L_\mathrm{BEV}=\frac{\ell_0+\frac12\ell_1+\frac14\ell_2}
 {1+\frac12+\frac14}.
 \label{eq:totalloss}
\end{equation}
Thus the scored patient-space objective remained primary while deep supervision
constrains the decoder at multiple BEV scales.

The submitted v2 checkpoint was trained from scratch for 27 epochs with the
corrected topological patient mask, using AdamW
(learning rate $5\times10^{-4}$, weight decay $10^{-5}$), polynomial decay
(power 0.9), BF16 autocast, unit-norm gradient clipping, seed 12345, and no
dropout or geometric augmentation. Ten randomly sampled beamlets per beam were
used in each update. An exponential moving average (EMA; decay 0.999) was
selected by validation high-dose MAE. The same checkpoint was used for abdominal and thoracic cases.

\paragraph{MRI-to-CT conversion.}
Training used
$48\times192\times192$ patches at $3\times1\times1\,\mathrm{mm}^3$ spacing,
batch size 2, MR Z-score normalization (input), and global CT-intensity normalization (output).
The regression objective used a deep-supervised MAE equivalent to the nnUNets default loss. The fold-all run was trained for 1000 epochs with Nesterov SGD
(initial learning rate $0.01$, momentum $0.99$, weight decay $3\times10^{-5}$)
and polynomial learning-rate decay, without mirroring or geometric
augmentation. At inference, overlapping $48\times192\times192$ windows are
combined with Gaussian weighting; the resulting sCT is passed unchanged to the
CT dose pipeline.

\subsection{Evaluation}
\label{sec:eval}
We use the official DoseRAD2026 metrics: Level~1 per-beamlet masked MAE (reference
dose above 10\% of that beamlet's maximum) and image-$z$ IDD distance, and
Level~2 stratified plan MAE, and local gamma pass rate (1\%/1\,mm).
Local validation values were first
averaged over beamlets within each case and then over the eight cases. The LUT
ablation used every fourth beam in every case, the same beam indices for all
LUTs, no learned correction, and the current evaluation code.

\section{Results}
\subsection{Water calibration}
In matched water benchmarks, the optimized $10^7$ LUT reached mean MAE
$0.26\%$ of peak and mean local gamma pass rate $98.8\%$ (1\%/1\,mm; 43
energies). The $10^8$ fit reduced mean MAE to $0.16\%$ and increased gamma to
$99.8\%$ over its initial 69-energy evaluation (worst MAE $0.32\%$, worst gamma
$98.2\%$); the 16 subsequently added energies remained in-family (MAE $0.14\%$,
gamma $99.6\%$). The Stage-2 engine-based fit particularly improves the distal
edge (e.g.\ $0.77\%\rightarrow0.20\%$ MAE at $164.45\,\mathrm{MeV}$ for the
earlier $10^7$ data). A $10^9$ refit converged cleanly but did not improve dose
accuracy on the patient validation set.

\begin{table}[H]
\centering
\caption{Matched LUT ablation on the eight validation cases. Values are
case means from the analytical engine without learned correction, using every
fourth beam and the topological mask; parentheses give the between-case standard
deviation. The $10^7$ LUT contains 43 fitted energies; the others contain 85.}
\label{tab:lut}
\small
\begin{tabular}{lcccc}
\toprule
Histories & $n_E$ & beam MAE & IDD distance & plan MAE\\
\midrule
$10^7$ & 43 & 0.01523 (0.00370) & 0.00765 (0.00109) & 0.03379 (0.00586)\\
$10^8$ (shipped) & 85 & \textbf{0.01347} (0.00346) & \textbf{0.00705} (0.00142) & \textbf{0.03140} (0.00590)\\
$10^9$ & 85 & 0.01348 (0.00348) & 0.00707 (0.00133) & 0.03142 (0.00589)\\
\bottomrule
\end{tabular}
\end{table}

Relative to the $10^7$-generation LUT, $10^8$ reduced mean beam MAE by
11.6\%, IDD distance by 7.9\%, and plan MAE by 7.1\%. This comparison includes
the increased fitted-energy coverage and removal of the unstable 142.06-MeV
entry, not only reduced MC noise. With identical 85-energy coverage, increasing
histories from $10^8$ to $10^9$ changed beam MAE by $+0.05\%$, IDD distance by
$+0.25\%$, and plan MAE by $+0.07\%$; signs were mixed across cases and the
difference was negligible. We therefore retained the $10^8$ LUT.

\subsection{Architecture selection}
Table~\ref{tab:conditioning} summarizes the controlled early conditioning
screen. Each configuration used the same small RepVGG--U-Net, data split,
optimizer, and 6,400 updates; only the listed energy representation,
spot-width input, and normalization were changed. The learned energy embedding
was the strongest representation: with GroupNorm and spot-width conditioning
fixed, its high-dose MAE was 7.5--7.6\% lower than the scalar and Fourier MLPs
and 10.8\% lower than omitting energy. Across the six matched normalization
pairs, GroupNorm was better in five and reduced their mean high-dose MAE from
1.831\% to 1.795\%. Because these short runs were not repeated exhaustively,
the normalization difference is interpreted as a selection result rather than
an estimate of generalization uncertainty.

\begin{table}[H]
\centering
\caption{Controlled conditioning screen after 6,400 updates (validation
high-dose MAE in percent; lower is better). All entries use the same seed and
validation beamlets. A dash denotes a configuration that was not run.}
\label{tab:conditioning}
\small
\begin{tabular}{llcc}
\toprule
Energy input & Spot MLP & Group & Instance\\
\midrule
Class embedding   & yes & \textbf{1.694} & 1.796\\
Class embedding   & no  & \textbf{1.683} & 1.775\\
Normalized scalar & yes & \textbf{1.832} & 1.872\\
Normalized scalar & no  & \textbf{1.848} & 1.892\\
Fourier features  & yes & 1.834 & \textbf{1.764}\\
Fourier features  & no  & \textbf{1.880} & 1.885\\
None              & yes & 1.899 & --\\
\bottomrule
\end{tabular}
\end{table}

Two additional within-seed comparisons supported the selected embedding. With
spot-width conditioning it reached 1.337\% and 1.093\%, compared with
1.547\% and 1.132\% without spot widths and 1.567\% and 1.166\% for scalar
conditioning. The validation beamlets differed between these seeds, so only
the ordering within each comparison is meaningful. Entrance addition and
AdaGN were effectively tied in a separate fixed-validation screen (1.472\%
versus 1.463\%), whereas combining both was worse (1.700\%); we retained the
simpler entrance injection.

We also screened the BEV deep-supervision weight under BF16. Weights 0, 0.05,
0.2, and 0.5 reached short-run high-dose MAE of 1.1545\%, 1.1108\%, 1.1053\%,
and 1.1388\%, respectively. Although 0.2 was marginally best numerically,
inspection of the predicted dose fields showed increasingly visible
grid-aligned structure at the larger auxiliary weights, consistent with the
coarser decoder targets exerting excessive influence. We therefore selected
0.05 as a compromise between the improvement over no deep supervision and
preservation of smooth native-resolution predictions. All four runs ended
after approximately 6,400--7,100 updates, so this screen supports the selected
low weight but does not establish a precisely optimal value. Earlier apparent
instability at high weights was traced to FP16 overflow and was not used for
this selection.

Beyond this controlled screen, longer depth kernels and the residual U-Net
progression improved validation accuracy, whereas attention modules, latent
depth mixing, combined entrance/AdaGN conditioning, stronger weight decay, and
geometric augmentation did not provide a reliable gain. These variants were
computationally expensive and were evaluated in short or single-run
experiments rather than a complete hyperparameter sweep. Scaling beyond the
887k model was similarly unproductive: at 17,728 matched updates, EMA
high-dose MAE was 0.963\%, 0.995\%, and 0.955\% for the 887k, 1.10M, and
1.45M models, respectively. The marginal 1.45M difference was not sustained in
a completed run, and no larger model completed the 64k or full-beam training
regime; the 887k model subsequently reached 0.802\% at 64k updates and 0.711\%
after full-beam training.

\subsection{Validation set (8 held-out patients)}
Table~\ref{tab:val} compares the two correction checkpoints under the same
$10^8$ engine and the pre-topological-mask local evaluation. Adding the
range-relative feature and training v2 from scratch improved all eight cases;
mean MAE decreased by 12.5\% and IDD distance by 5.6\% relative to v1 in a
separate every-fourth-beam paired sweep. In the full evaluation, v2 reached
MAE $0.00571$ and IDD distance $0.00463$. Its integral ratio of $0.988$ denotes
a small residual under-response rather than exact dose conservation. Errors
remained larger in thoracic than abdominal cases.

\begin{table}[H]
\centering
\caption{Full local validation set before the topological-mask update
(mean over eight cases). Integral ratio is a local diagnostic.}
\label{tab:val}
\begin{tabular}{lccc}
\toprule
Model (engine identical; MCS on) & Beam MAE & IDD & Integral\\
\midrule
v1 additive corrector (8 channels; 13 epochs) & 0.00574 & 0.00475 & 0.988\\
\textbf{v2 additive corrector (9 channels; 17 epochs)} & \textbf{0.00571} & \textbf{0.00463} & 0.988\\
\bottomrule
\end{tabular}

\end{table}

\paragraph{Corrected mask and extended training.}
The 13-epoch submission was trained with the density-threshold mask and only
evaluated with the topological one. We retrained the same recipe from scratch
with the corrected mask and a longer schedule. Table~\ref{tab:epochs} follows
that run on the eight held-out cases: at 27 epochs beam MAE falls by 5.12\% and
IDD distance by 8.99\%, improving both metrics on all eight cases.

\begin{table}[H]
\centering
\caption{Validation set across training. All rows use the topological mask,
every sixth beam, per-beamlet Level~1 metrics, EMA weights, and an identical
engine and LUT; only the checkpoint differs. Mean over the eight cases with the
between-case standard deviation in parentheses; percentages are relative to the
submitted 13-epoch checkpoint.}
\label{tab:epochs}
\small
\begin{tabular}{lcccc}
\toprule
Checkpoint & Beam MAE $\times 10^{-3}$ & $\Delta$ & IDD $\times 10^{-3}$ & $\Delta$\\
\midrule
13 epochs (submitted; density mask) & 5.37 (0.88) & --- & 4.29 (1.41) & ---\\
14 epochs & 5.31 (0.86) & $-0.95\%$ & 4.23 (1.42) & $-1.40\%$\\
17 epochs & 5.25 (0.82) & $-2.13\%$ & 4.16 (1.39) & $-3.03\%$\\
20 epochs & 5.19 (0.81) & $-3.25\%$ & 4.33 (1.39) & $+0.92\%$\\
23 epochs & 5.14 (0.79) & $-4.20\%$ & 4.42 (1.43) & $+3.15\%$\\
24 epochs & 5.13 (0.80) & $-4.32\%$ & 4.08 (1.38) & $-4.83\%$\\
26 epochs & 5.10 (0.78) & $-4.91\%$ & 4.05 (1.37) & $-5.53\%$\\
\textbf{27 epochs} & \textbf{5.09 (0.78)} & $\mathbf{-5.12\%}$ & \textbf{3.90 (1.36)} & $\mathbf{-8.99\%}$\\
\bottomrule
\end{tabular}
\end{table}

\subsection{Preliminary hidden-test results}
Table~\ref{tab:test} reports the platform evaluations of 12 and 24 August 2026.
The mask correction was the largest late-stage change: for the local THB016
beam~31 diagnostic it reduced MAE from 0.01568 to 0.00872 and IDD distance from
0.01948 to 0.00481 by retaining dose in internal air cavities. Retraining moved
the two tracks very differently: on CT, beam MAE improved by 3.0\% and IDD
distance by 8.0\%, whereas on MRI beam MAE improved by 0.5\% and IDD distance
not at all. The MRI track uses the same dose model after synthetic-CT
generation, so this confirms the MR-to-CT front end, rather than the dose
engine, as the dominant remaining limitation there.

\begin{table}[H]
\centering
\caption{Preliminary hidden-test metrics. All submissions apply the topological
patient mask at inference.}
\label{tab:test}
\small
\begin{tabular}{llccccc}
\toprule
Track & Checkpoint & Beam & IDD & Plan & Gamma (\%) & DVH\\
\midrule
\textit{before fixing mask} &&&&&\\

CT & 13 epochs & 0.0100 & 0.0094 & 0.0092 & 97.20 & 0.5955\\
\midrule
\textit{after fixing mask} &&&&&\\
CT & 13 epochs & 0.0066 & 0.0025 & 0.0049 & 98.30 & 0.460\\
CT & 17 epochs & 0.0064 & 0.0023 & 0.0047 & 98.51 & 0.448\\
CT & 27 epochs & \textbf{0.0061} & \textbf{0.0020} & \textbf{0.0044} & \textbf{98.81} & \textbf{0.431}\\
\midrule
\textit{before fixing mask} &&&&&\\
MRI & 13 epochs & 0.0189 & 0.0154 & 0.0188 & 84.03 & 3.771\\
\midrule
\textit{after fixing mask} &&&&&\\
MRI & 13 epochs & 0.0185 & 0.0152 & 0.0185 & 84.02 & 3.853\\
MRI & 17 epochs & 0.0184 & 0.0152 & 0.0184 & 84.10 & 3.813\\
MRI & 27 epochs & \textbf{0.0183} & \textbf{0.0151} & \textbf{0.0183} & \textbf{84.15} & \textbf{3.806}\\
\bottomrule
\end{tabular}
\end{table}

\subsection{Runtime}
An earlier CT platform submission required $135\,\mathrm{s}$ per canonical
case; MRI submission required $144.3\,\mathrm{s}$. To increase the speed we switched our models and engine to float16 instead of float32, activated RepVGG inference fusion, increased batching for engine (b=128), for correction model (b=8) and also image writing was optimised (threaded image writing).
However, the algorithm was still
implemented in PyTorch and Python so substantial speed improvements are still 
possible. 

\section{Discussion}
\label{sec:discussion}
The comparison in Table~\ref{tab:lut} shows diminishing returns from increasing
water-phantom histories: $10^8$ removes the instability and noise present in the
$10^7$ calibration, whereas $10^9$ does not measurably improve patient dose.
Accuracy is instead driven by the learned correction and by correct anatomical
support. In particular, the topological mask explains much of the late IDD and
plan-level gain, so the remaining error cannot be attributed solely to the
lateral kernel.

The additive, zero-initialized residual provides a stable physics baseline, and
the BEV representation aligns the narrow Bragg peak across gantry angles. The
material and energy embeddings complement continuous SPR/WED features,
while multi-scale BEV supervision regularizes intermediate decoder resolutions
without replacing the official patient-space target. A residual far-halo
under-response remains, especially at high energy. Increasing the lateral fit
window alone did not solve it and could overestimate halo dose, indicating a
shape/support trade-off in the double-Gaussian model rather than simple spatial
truncation.

Differentiability is an additional design property of the physics engine rather
than a requirement of the DoseRAD task. It makes the forward model reusable as
a component of gradient-based treatment-planning workflows, where dose-derived
objectives can be optimized through the engine. The present challenge results
demonstrate differentiable kernel calibration, but do not constitute an
evaluation of plan optimization.

Limitations include only eight cases for local model selection, tuning and
reporting on the same validation cohort, and incomplete runtime provenance for
the final CT artifact. Checkpoint selection by validation high-dose MAE is
insensitive to IDD distance, which is not monotonic during training
(Table~\ref{tab:epochs}); a criterion combining both scored quantities would be
preferable. The MRI results additionally depend on synthetic-CT quality.

\subsubsection{Author contributions (CRediT).}

Lukas Zimmermann -- Conceptualization, Data curation, Formal analysis, Investigation, Methodology, Resources, Software, Validation, Visualization, Writing – original draft.
Hermann Fuchs -- Resources, Data curation, Writing – review \& editing.
Attila  Simk\'o -- Investigation, Software, Writing – review \& editing.
Gerd Heilemann -- Project administration, Writing – review \& editing.

\subsubsection{Code availability.}
The proton dose engine is part of the open-source PyDoseRT
package.\footnote{\url{https://github.com/UMU-DDI/PyDoseRT}} The physics engine,
challenge container, training code, evaluation scripts, and model definitions
will be available in a separate repository.

%


\begin{thebibliography}{8}
\bibitem{ref_doserad}
Xiao, F., Delopoulos, N., Wahl, N., et al.: DoseRAD2026 Challenge dataset:
AI accelerated photon and proton dose calculation for radiotherapy.
arXiv:2604.12778 (2026)

\bibitem{ref_pydose}
Simk\'o, A., Kronsteiner, M., Glatzer, S., et al.: A physics-informed,
plug-and-play dose engine for gradient-based radiotherapy treatment planning.
arXiv:2512.18863 (2025)

\bibitem{ref_gate}
Sarrut, D., et al.: The OpenGATE ecosystem for Monte Carlo simulation in
medical physics. Phys. Med. Biol. (2022)

\bibitem{ref_hong}
Hong, L., Goitein, M., Bucciolini, M., et al.: A pencil beam algorithm for
proton dose calculations. Phys. Med. Biol. \textbf{41}, 1305--1330 (1996)

\bibitem{ref_kanematsu}
Kanematsu, N.: Alternative scattering power for Gaussian beam models of charged
particles. Nucl. Instrum. Methods B \textbf{266}, 5056--5062 (2008)

\bibitem{ref_repvgg}
Ding, X., Zhang, X., Ma, N., et al.: RepVGG: making VGG-style ConvNets great
again. In: CVPR (2021)

\bibitem{ref_nnunet}
Isensee, F., Jaeger, P.F., Kohl, S.A.A., Petersen, J., Maier-Hein, K.H.:
nnU-Net: a self-configuring method for deep learning-based biomedical image
segmentation. Nat. Methods \textbf{18}, 203--211 (2021)

\bibitem{ref_impact}
Boussot, V., H\'emon, C., Nunes, J.-C., et al.: IMPACT: a generic semantic loss
for multimodal medical image registration. arXiv:2503.24121 (2025)

\bibitem{ref_Fuchs2012}
H.~Fuchs, J.~Ströbele, T.~Schreiner, A.~Hirtl, and D.~Georg,
``A pencil beam algorithm for helium ion beam therapy,''
Medical Physics, \textbf{39(11)}, p.~6726, (2012).
doi: 10.1118/1.4757578.


\end{thebibliography}
\end{document}